\documentclass[aps,final,a4paper,twocolumn,prl]{revtex4-1}
\usepackage{amsfonts}
\usepackage{amsmath}
\usepackage{amssymb}
\usepackage{graphicx}

\begin{document}
\title{Theory of Tunneling Spectroscopy in a Mn$_{12}$ 
Single-Electron Transistor by Density-Functional Theory Methods}
\author{ \L. Michalak$^{1}$, C. M. Canali$^{1}$}

\author{M. R. Pederson$^{2}$}

\author{M. Paulsson$^{1}$}

\author{V. G. Benza$^{3}$}

\affiliation{$^{1}$Division of Physics, Department of Natural Sciences, Kalmar
University, 391 82 Kalmar, Sweden}

\affiliation{$^{2}$Center for Computational Materials Science, Naval Research Lab, Code 6390, Washington, DC 20375, USA}

\affiliation{$^{3}$Dipartimento di Fisica e Matematica Universit\'a dell' Insubria, 20064 Como, Italy}



\begin{abstract}
We consider tunneling transport through a Mn$_{12}$ molecular magnet
using spin density functional theory. 
A tractable methodology for constructing many-body wavefunctions from Kohn-Sham 
orbitals allows for the determination of spin-dependent matrix elements 
for use in transport calculations.
The tunneling conductance at finite bias is characterized by
peaks representing transitions between spin multiplets, 
separated by an energy on
the order of the magnetic anisotropy.
The energy splitting of the spin multiplets and 
the spatial part of their many-body wave functions,
describing the orbital degrees of freedom of the excess charge,
strongly affect the electronic transport, and can lead to 
negative differential conductance.
\end{abstract}
\volumeyear{2010}
\volumenumber{104}
\eid{10.1103/PhysRevLett.104.017202}
\received[Received ]{10 November 2008}

\revised[revised manuscript received ]{2 November 2009}


\published[published ]{5 January 2010}

\maketitle

There is a growing interest in exploring the rich
physics and spintronics functionality of molecular
single-electron transistors (SETs) consisting of a few {\it magnetic} 
molecules
weakly coupled to nano-gapped electrodes \cite{wernsdorfer_nature08}.
Recently two groups \cite{heersche_prl06,mhjo_nano06} 
have carried out single-electron
tunneling experiments on individual magnetic molecules 
based on Mn$_{12}$O$_{12}$ (henceforth Mn$_{12}$) with organic ligands.
Mn$_{12}$ is the most studied and perhaps the most remarkable
molecular magnet \cite{molmagnet}. 
In its crystal phase, Mn$_{12}$ is characterized by a long 
spin relaxation time due to its large uniaxial magnetic-anisotropy energy.
Furthermore, at low temperatures, quantum effects in the relaxation 
properties are clearly 
discernible \cite{sessoli1993, thomas1996,friedman1996,
wernsdorfer1999_science} 
and have been attributed to 
quantum tunneling of 
the molecule collective 
magnetization \cite{molmagnet}.
How these properties are revealed in electronic 
quantum transport is
a question of great significance for the field of molecular 
spintronics \cite{wernsdorfer_nature08}. 
Indeed, the SET experiments \cite{heersche_prl06,mhjo_nano06}
show signatures of the molecule 
magnetic state and its low-energy collective spin excitations. 
The theoretical models proposed so far 
\cite{heersche_prl06,mhjo_nano06,romeike_prl2006b,timm2006prb,mucciolo2006prl,
timm2007prb,elste2007prb,romeike_prb2007,mucciolo2008prb}, 
are typically based on
effective giant-spin Hamiltonians with large uniaxial anisotropy barriers.
This approach has two drawbacks \cite{lehmann_prl2007}. 
First, the effective
spin Hamiltonian for the charged states (anion and cation) of Mn$_{12}$
needed to describe sequential tunneling transport, 
is not known.  
Scaling of
the global
anisotropy parameter to account empirically for changes in the number of
electrons
forming a macro-spin is fraught with uncertainty \cite{park2003prb}.
Second, 
the orbital degrees of freedom 
are not included in the giant-spin Hamiltonian formalism. 
The orbital effects due to changes in electron population on the 
Mn$_{12}$ molecule 
modify the symmetry
and magnitude of the magnetic-anisotropy Hamiltonian and can even
change the spin ordering~\cite{park2004prb}. 

In this Letter we provide a microscopic many-body 
description of the ground
state (GS) and low-lying spin excitations of both neutral and {\it charged}
states of a Mn$_{12}$ molecular magnet. Our approach is based on 
spin density functional
theory (SDFT), which has been very successful in describing the 
spin-orbit-induced magnetic anisotropy barrier in Mn$_{12}$ 
and other molecular
magnets \cite{pederson_prb99, pederson2000stsol, 
postnikov2006polyhedron}. 
We find that 
when a
{\it delocalized} electron 
is added to (subtracted from) the molecule, 
the GS spin of the 
molecule increases (decreases) by 1/2. 
For both charged states,
the GS magnetic-anisotropy energy
is larger than for the neutral Mn$_{12}$.
We then incorporate this information 
into a quantum master equation for electronic transport in the 
sequential tunneling approximation,
which is appropriate for the experimental Coulomb
blockade (CB) regime. 
The approximate many-particle eigenstates 
lead to a 
tunneling conductance that exhibits fine structure on the order
of the anisotropy energy and, under certain circumstances, 
to strong negative differential conductance
(NDC).
Comparison with the giant-spin model 
%
shows that spatial selection rules play a crucial role in determining
which spin excitations contribute the most to the
tunneling conductance.

We need to know the many-electron wave functions, representing low-energy
spin excitations, as a 
function of 
the excess charge ($Q$), 
spin ordering ($\mathbf{M} \equiv \{\mu_{\nu}\}_{\nu =1,12}$), 
applied electric $\mathbf{E}$ and magnetic $\mathbf{B}$ fields, and 
the parameters $\theta$, $\phi$ describing the quantization axes. 
We refer to the collection of all possible variables as the 
"order-parameter vector" (OPV), 
$\mathbf{p} \equiv (Q, \mathbf{M}, \mathbf{E}, \mathbf{B}, \theta, \phi)$ 
to label the states. Given a specific OPV, we first construct a set of 
Kohn-Sham (KS) single-particle states $\Phi_k(\mathbf{p})$, by diagonalizing a KS single-particle Hamiltonian 
$H(\mathbf{p})$ that depends upon this OPV. 
Since some of these effects ($Q, \mathbf{M}$) 
are clearly large and some 
($\theta, \phi$) are 
generally small, there is flexibility as to 
which of these terms 
must be accounted for self-consistently.
Specifically,
\begin{eqnarray}
H(\mathbf{p}) &=& H(Q,\mathbf{M},\mathbf{E},\mathbf{B},\theta,\phi) 
\nonumber \\
&=& H_0({\rm DFT},Q,\mathbf{M}) + V_{\mathbf{L} \cdot \mathbf{S}}(\theta, \phi)
+ \mathbf{E} \cdot \mathbf{r} 
\nonumber \\
&& + \mathbf{B} \cdot (\mathbf{L} + 2\mathbf{S}),
\label{ham}
\end{eqnarray}
\noindent contains a spin-polarized term $H_0({\rm DFT},Q,\mathbf{M})$, which is treated self-consistently
for the cation, neutral and anionic 
states ($Q = +1, 0$ and $-1$); 
$V_{\mathbf{L} \cdot \mathbf{S}}(\theta, \phi)$ represents the spin-orbit
interaction. 
We neglect the last two terms
representing the coupling to external fields.

The spin-ordering $\mathbf{M}$ corresponds to that obtained from 
the local moments of the 12 Mn atoms ($\mu_{\nu}$) in the classical 
ferrimagnetic 
state of the neutral molecule \cite{kortus-v15_prl2001}. 
The spin-orbit operator is treated exactly~\cite{pederson_prb99} \--- 
albeit non-self-consistently \--- 
in the basis of the eigenstates of $H_0({\rm DFT},Q,\mathbf{M})$.
Diagonalizing the above Hamiltonian with the constraint that the expectation value of the total 
spin ($\langle \mathbf{S} \rangle$) is quantized along the axis determined by $\theta, \phi$ results 
in a set of single-particle, {\it noncollinear} spin orbitals, $\phi_k(\mathbf{p})$, expressed 
as $\phi_k = \phi^+_k(r) \chi_+(\theta,\phi) + \phi^-_k(r) \chi_-(\theta,\phi)$. 
Here, the angle-dependent spinors $\chi_{\pm}(\theta,\phi)$ are spin-1/2
coherent states specified by the quantization axis, 
$\chi_{\pm}(\theta,\phi) = \cos(\theta/2)|\pm \rangle \pm \sin(\theta/2)e^{\pm i \phi}|\mp \rangle$.

\begin{figure}
  \includegraphics[height = 4.6 cm]{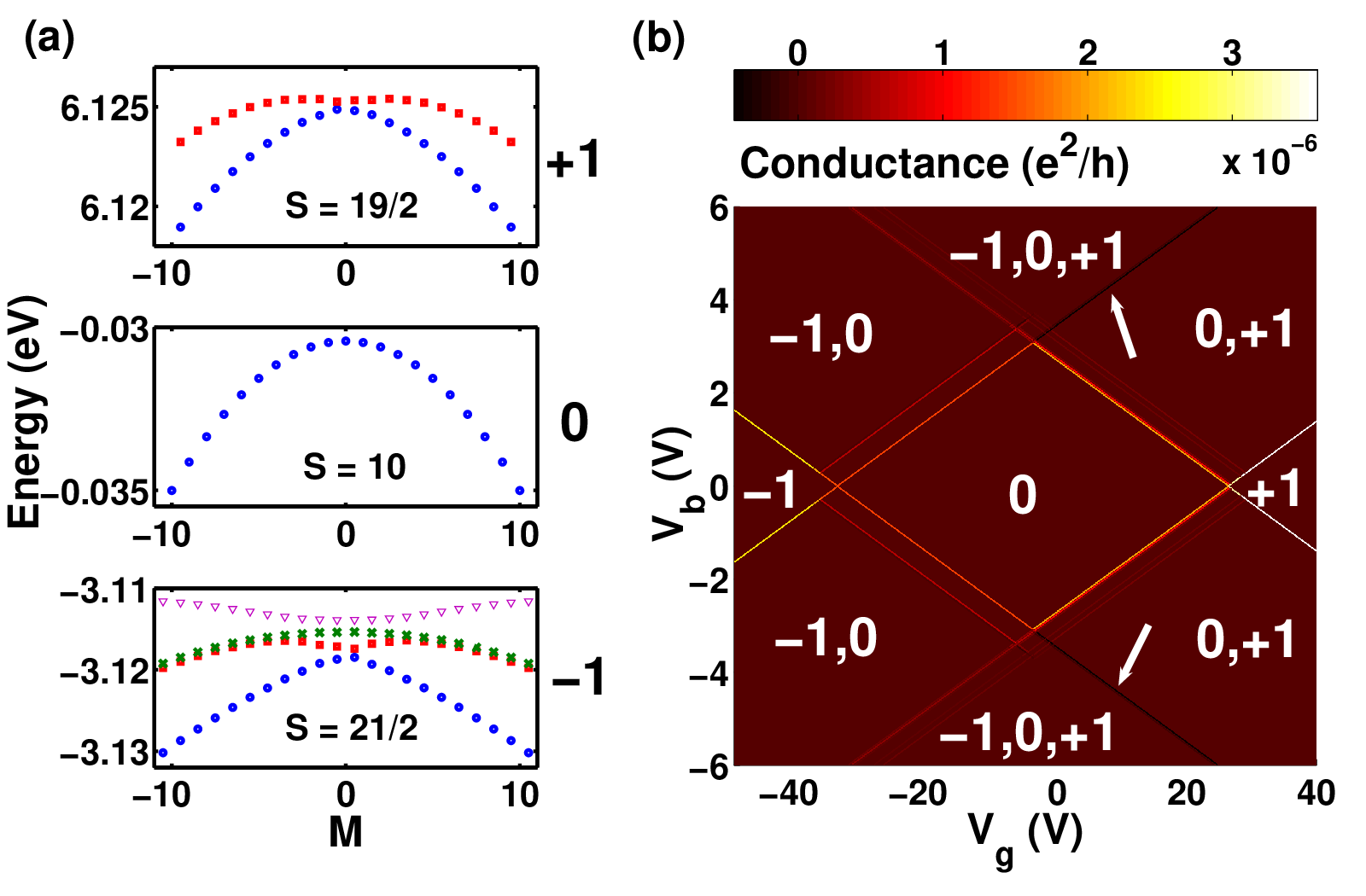}
  \caption{(color online).
(a) Lowest lying spin multiplets for the three charge states of the molecule; 
the small
energy difference between states $M$ and $-M$ is due to transverse anisotropy terms 
coming from 4th-order spin-orbit contributions and our choice of the $2S+1$ values of 
($\theta_M$, $\phi_M$). See text after Eq.~(\ref{slater}).
(b) Differential conductance of a Mn$_{12}$ SET as a function of bias and gate voltage, 
for a symmetric double junction, 
with gate capacitance equal to 1/20 of the total capacitance, at zero temperature. 
Numbers -1, 0 and +1 denote the excess charge on the molecule and label regions of Coulomb blockade. 
In regions $(-1,0)$, $(0,+1)$ and $(-1,0,+1)$ transport is possible via transitions between different charge states.}
  \label{dc2D}
\end{figure}

We now construct approximate many-body functions for the ground and excited electronic states as
single Slater determinants (SDs) of the spin orbitals $\phi_k(p)$:
\begin{equation}
\vert \mathbf{p}; k_1, k_2,\dots k_{N_Q} \rangle 
\equiv \vert \phi_{k_1}(\mathbf{p})\phi_{k_2}(\mathbf{p}) \dots \phi_{k_{N_Q}}(\mathbf{p}) \rangle.
\label{slater}
\end{equation}
The above states are generally not eigenstates 
of either ${\bf S}^2$ or $S_z$. However, 
a state with $\vert \langle {\bf S}^2 \rangle \vert = S_0(S_0+1)$, especially when constructed from a closed shell 
of spatial states, is expected to be the primary contributor to an $S = S_0$ eigenstate.
While the variables ($\theta, \phi$) generate a continuous overdetermined set of SDs, 
a judicious choice of $2S_0+1$ values of $\theta$ and $\phi$ can
lead to a nearly orthogonal set of normalized linearly independent many-particle SDs,
with $\langle S_z \rangle$ taking on integer or half integer values 
akin to the standard $M= -S_0, -S_0+1, \dots, S_0$ quantum numbers.
Choosing $2S+1$ values of $\theta$ given by $S\cos(\theta_M)=M$ leads to 
integer or half integer $\langle S_z \rangle$
regardless of the choice of $\phi$; 
however, choosing $\phi_M = M \phi_o$
leads to destructive interference in the off-diagonal elements of these states and aids in producing
approximate $S_z$ eigenstates. For the case of $S=10$, we find that choosing $\phi_o=34^o$ leads to the
smallest off-diagonal overlaps between approximate eigenstates. 
%
We call the $2S_0+1$ many-electron states constructed with this procedure a spin multiplet. 
Besides the GS spin multiplet, the anion and cation
have a few low-lying excited spin multiplets. These come about because
the HOMO level of the charged molecule is quasidegenerate with 
a many-fold of LUMO levels \cite{park2004prb}. Using Eq.~(\ref{slater}), we can construct several
SD excited-states close in energy to the GS, all having the same spin, $S_0$.
The relevant spin-multiplets for the molecule are shown in Fig.~\ref{dc2D}(a).
Typically, the level spacing within each spin multiplet is of the order of
0.1-1 meV, while different multiplets are separated by energies of the order of 10 meV. 
Note that 
the energies of a given spin multiplet
are not exactly invariant under
$M \rightarrow -M$, since the choice of the angle $\phi_M$ 
is incommensurate with the 
nonperfectly uniaxial symmetry
of Mn$_{12}$.
The breaking of the level degeneracy for the $(M, -M)$ pair is
of the order of the transverse anisotropy terms coming 
from 4th order spin-orbit 
contributions, and therefore very small \cite{note1}.
For a later comparison with the giant-spin approach, we disregard the small deviations 
from uniaxial symmetry and consider  
\begin{center}
\begin{table}[h]
    \caption{The GS properties from DFT: spin, 
energy and magnetic-anisotropy energy as a function of charge.}
    \begin{tabular}{ | l | c | c | c| c | c |}
    \hline
    State & $Q$ & Spin & Energy (eV) & MAE (K) & MAE (meV)\\ \hline
    Anion & -1 & 21/2 & -3.08 & 137 & 11.8\\ 
    Neutral & 0 & 20/2 & 0.00 & 55 & 4.7\\ 
    Cation & 1 & 19/2 & 6.16 & 69 & 5.9\\
    \hline
    \end{tabular}
\label{table}
\end{table}
\end{center}
the spin Hamiltonian 
$H_{S_Q} = \sum_{i,n} D_{S_Q, i, n} 
\big[S^z_{Q, i}\big]^n$,
given in terms of spin variables
$\vec{S}_ {Q, i}$ for each spin-multiplet
${i}$ of charge state $Q$;
$D_{S_Q, i, n}$ are anisotropy constants which we extract by fitting
the corresponding SDFT energy spectra.
Table~\ref{table} shows
the total spin, GS energy and magnetic-anisotropy energy of the lowest
spin multiplet calculated
for the different charge states of Mn$_{12}$.
Adding
an electron to the molecule increases the molecule 
spin and decreases the energy.
Furthermore,
the anisotropy increases significantly when a 
delocalized electron is added to the neutral molecule. 
This is due to the fact that there is a near degeneracy between unoccupied 
onefold and twofold states. 
The spin-orbit interaction leads to
a strong mixing 
between these states which, because of the orbital components, enhances the anisotropy.


In the following we discuss quantum transport through a Mn$_{12}$ molecule weakly coupled to metallic
electrodes. 
Electron tunneling between leads and the molecule is described 
by the Hamiltonian
$H_{\rm T} = \sum_{\alpha, l} \sum_{k, \mathbf{p}} t_{\alpha l} a^{\dagger}_{\alpha l} c_{k}(\mathbf{p}) + {\rm H.c.}$,
where $c^{\dagger}_{k}$ ($c_{k}$) creates (destroys) an electron in orbital $k$ in the molecule, 
and $t_{\alpha}$ depends on the tunneling barrier between lead $\alpha = {\rm L, R}$ (left or right) and the molecule. 
The leads, at electrochemical potential $\mu_{\alpha}$, are described by the independent-electron 
Hamiltonian $H_{\alpha} = \sum_{l}\epsilon_{\alpha l} a^{\dagger}_{\alpha l} a^{\phantom{\dagger}}_{\alpha l}$, 
where $a^{\dagger}_{\alpha l}$ ($a^{\phantom{\dagger}}_{\alpha l}$)
creates (destroys) 
a quasiparticle of quantum number $l$. 
We also take into account the work function for the external leads. 
As a result, the charge state populated at zero gate and bias voltages is the neutral one, 
and not the anion 
as it would seem from Table I.
The sequential tunneling current is obtained from a master equation for the occupation probabilities of the  
molecule many-body states.
The transition rate
between two many-body states
via tunneling of an electron from lead ${\alpha}$ into
the molecule, is proportional to
$f\big(E(\mathbf{p'}, k_i', N_{Q'}) - E(\mathbf{p}, k_i, N_{Q})- 
\mu_{\alpha}\big)\times
|\langle \mathbf{p'};k_1' ,k_2',\dots,k_{N_{Q'}}' \vert c_k^{\dagger}(\mathbf{p}) 
\vert \mathbf{p};k_1,k_2,\dots,k_{N_{Q}} \rangle|^2$,
where $f(E)$ is the Fermi distribution function
and $E(\mathbf{p}, k_i, N_{Q})$ is the energy of state $\vert \mathbf{p};k_1,k_2,\dots,k_{N_{Q}}\rangle$, modified by the
bias $V_{b} = (\mu_{L} - \mu_{R})/e$ and gate voltage $V_{g}$.
We then solve numerically the master equation in steady-state
and obtain the current as a function of $V_{b}$ and $V_{g}$.
Figure ~\ref{dc2D}(b) shows the differential conductance $G= dI/dV_{b}$ as function of  $V_{b}$ and $V_{g}$. 
The 
calculations are done at zero temperature. 
We choose equal coupling of the molecule to the two leads;
the gate capacitance is equal to 1/20 of the total capacitance of the system.   
Three CB stability diamonds 
are visible, corresponding to the three 
different charge states $Q= -1, 0, +1$, where transport is blocked.  
In region indicated by $(-1,0)$ [respectively $(0,+1)$],
current flows through transitions between 
anionic (cationic) and neutral states. 
In region $(-1,0,+1)$ all 
three charge states are present.
The additional lines parallel to the GS-GS transitions are due to transitions between excited states. 
%
%
In Fig. \ref{dc2D}(b) we can also see two lines, indicated by arrows,
that correspond to a decrease in the current with increasing $V_{b}$ (NDC). 
These lines give the bias at which, for a given $V_{g}$, anionic states become occupied 
in 
the $(0,+1)$ region.
NDC in Mn$_{12}$-SET has been 
observed experimentally ~\cite{heersche_prl06}.

\begin{figure}
  \includegraphics[height = 4.5 cm, width = 8 cm]{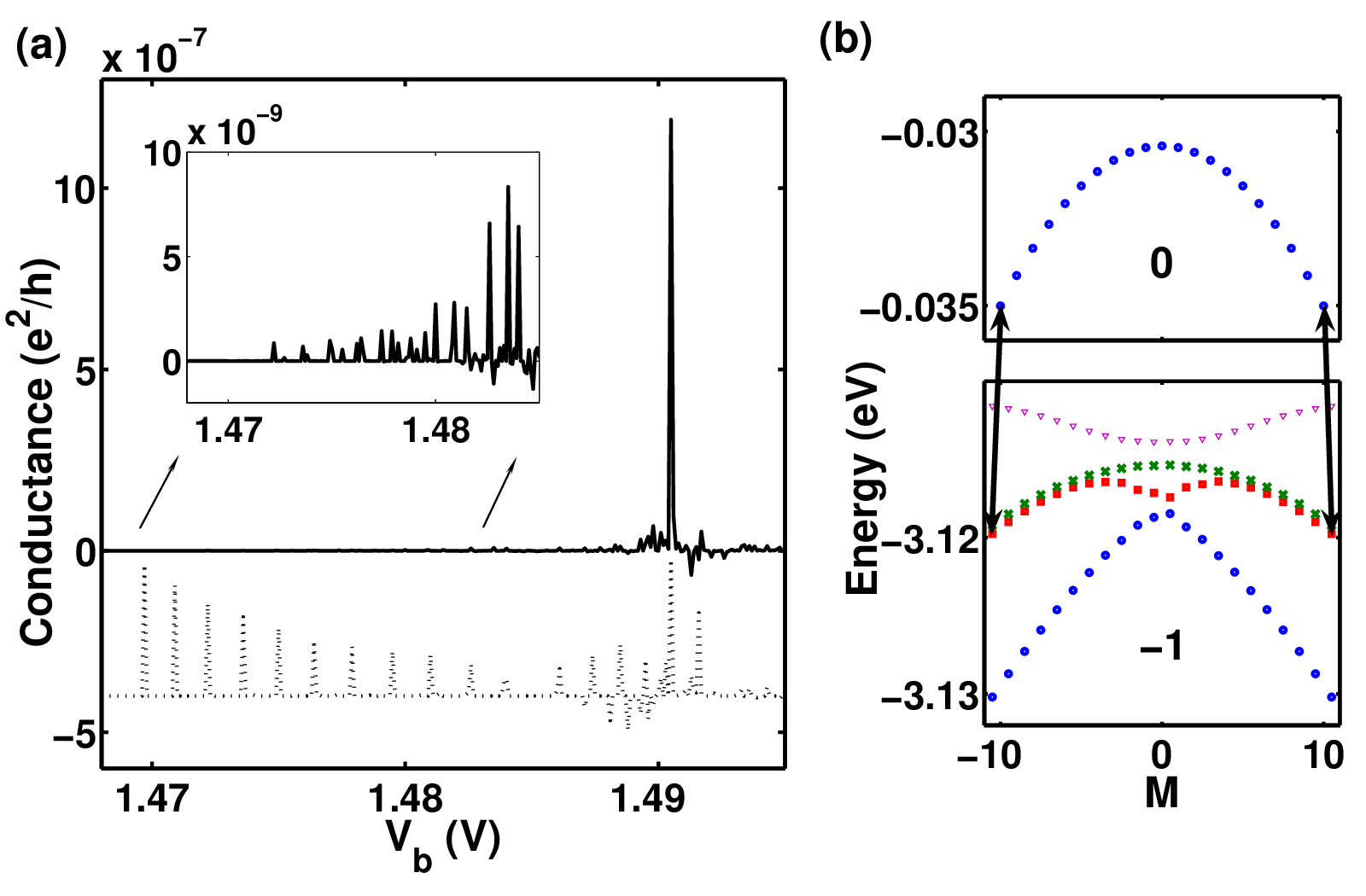}
\caption{(color online). 
(a)  Differential conductance as a function of bias at $V_g = -20$~V.
Parameters are as in Fig.~\ref{dc2D}.
Solid (dotted, offset for clarity) lines: calculation based on SDFT (giant-spin model).
Inset of (a) shows a zooming of the onset of SDFT transport
due to transitions between
the ground-state (GS) spin multiplets of the neutral and anionic molecule.
Visible peaks in the main plot correspond to transitions between
the GS spin multiplet of the neutral and the
first three excited multiplets
of the anion.
(b) Spin multiplets involved in the transport.
The transitions between the states
generating the dominant peak in (a) are indicated by arrows joining the states.}
  \label{step1}
\end{figure}

For a better understanding of transport just above the CB gap, 
in Fig.~\ref{step1}(a) we plot the differential conductance 
as a function of $V_{b}$, 
for $V_g = -20$V.
Transport in this region is due to 
transitions between the spin multiplets of the neutral and anionic molecule.
The conductance peak spectrum displays a rich fine structure, with peak spacing on the order of 0.1-1 meV, 
which corresponds to that seen in experiment~\cite{heersche_prl06, mhjo_nano06}. 
The first set of peaks at the very onset of transport is 
caused by transitions between the GS spin multiplets.
Surprisingly, the conductance in this region is very small, $ G \le 10^{-8} e^2/h$, as shown in the inset; 
it is practically invisible for transitions between the low-lying states (large $|M|$ and $|M'|$) and slightly
larger for transitions between higher-lying states (small $|M|$ and $|M'|$).
As we argue below,
this is caused by the very small overlap of the {\it orbital} parts of the many-body wave-functions 
of the two GS spin multiplets.
The second set of peaks in  Fig.~\ref{step1}(a), $V_{b} \ge 1.485$ V, corresponds to transitions between the GS spin
multiplet of the neutral molecule and the first three excited spin multiplets of the anion. 
This cluster of resonances is largely determined by the first excited spin multiplet of the anion,
since reaching this multiplet opens up transport also via other multiplets.
In particular, the dominant peak seen in the figure is due to transitions between 
the lowest-energy states of the GS spin multiplet of the neutral molecule 
and the first excited spin multiplet of the anion,
as shown in Fig.~\ref{step1}(b).
 
In order to 
shed light on the interplay between 
orbital and spin-selection rules, 
we compare the SDFT-based calculation
with the giant-spin model.
Within this spin model, transitions are possible only between states whose spin 
differs by 1/2 (spin-selection rule), 
with transition rates given by Clebsch-Gordan (CG) coefficients \cite{cmc_ahm2000prl}. 
In the computation of the conductance, we include the GS spin multiplet of the neutral molecule, 
and the GS and first three excited spinmultiplets of the anion. 
The conductance for the giant-spin model is shown with the dotted line in Fig.~\ref{step1}(a). 
The first 11 peaks in the inset correspond to subsequent transitions between states 
$M = \pm S, \pm (S-1), \ldots, 0$ and $M' = \pm S', \pm(S'-1), \ldots, 1/2$, where $S'= S+1/2$.
The intensity of these peaks decreases monotonically with decreasing $|M|$, 
which is different from the SDFT-based conductance.
The more complicated set of peaks at $V_{b} \ge 1.485$ V in Fig.~\ref{step1}(a) 
resembles the analogous cluster of peaks for SDFT and has the same interpretation.
\begin{figure}
  \includegraphics[height = 4 cm]{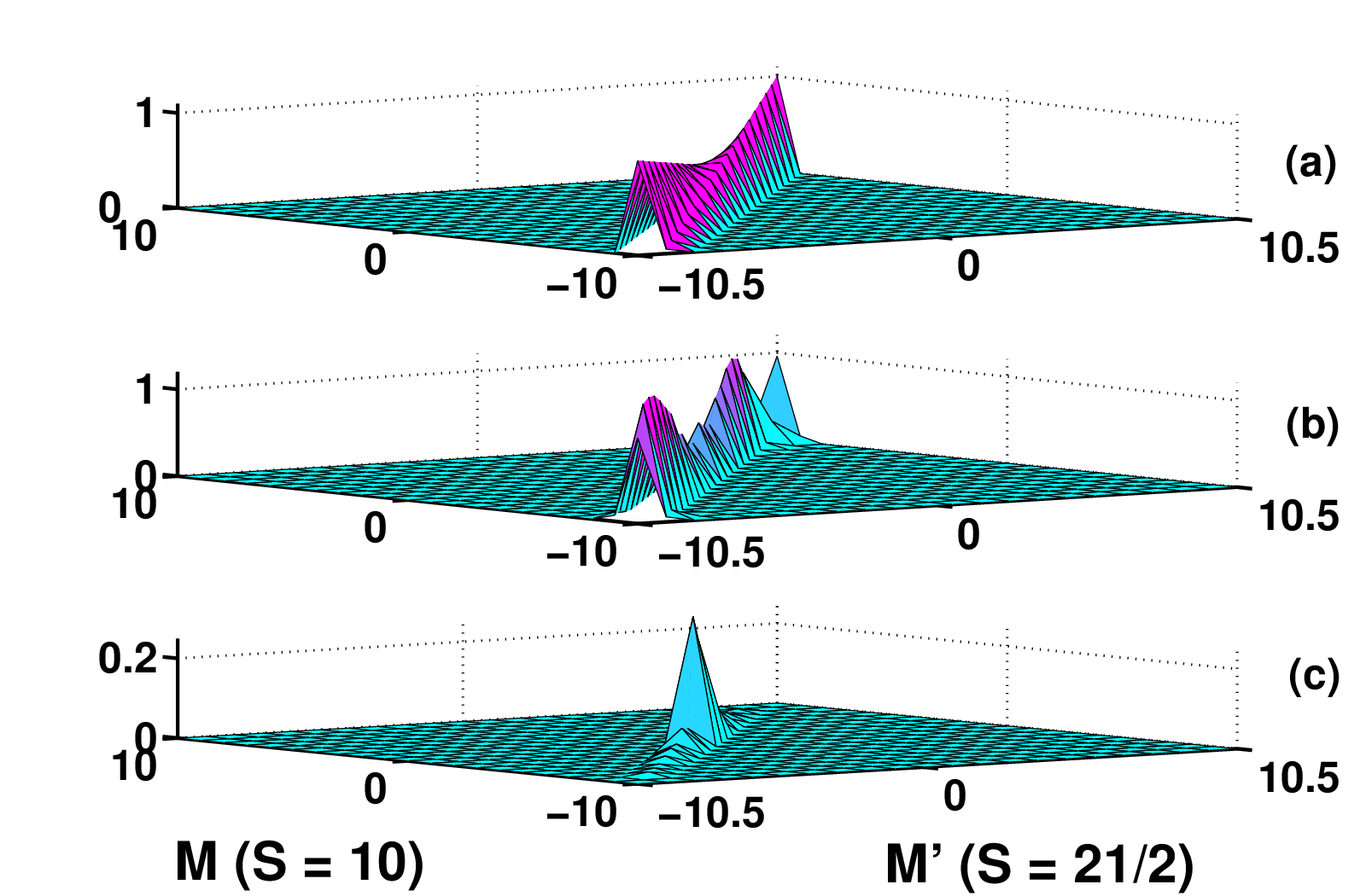}
  \caption{(color online). Matrix elements for transitions between anionic and neutral charge states. (a) Giant-spin model;
(b) SDFT results for transitions from neutral GS spin multiplet to anionic first excited multiplet; 
(c) SDFT results for transitions from neutral GS to anionic GS spin multiplets.} 
  \label{plot_overlaps}
\end{figure}
We examine the matrix
elements giving the neutral-to-anion transition rates. Figure~\ref{plot_overlaps}(a) shows the results
for the giant-spin model, where the matrix elements are proportional to the CG coefficients and
the spin selection rule $|M-M'| = 1/2$ is strictly obeyed.
Figure~\ref{plot_overlaps}(b) shows that the same spin-selection rule is 
approximately satisfied by the SDFT matrix elements of the transitions between the GS spin multiplet of
the neutral molecule and the first excited spin multiplet
of the anion. 
In particular, the orbital part of the wave functions does not modify substantially
this condition. 
In contrast, Fig.~\ref{plot_overlaps}~(c) shows that
the SDFT matrix elements for the transition between the two GS spin multiplets are 
different: the effect of the spin-selection rules is now overridden 
by space selection rules, which suppress most of the transition rates near $|M| = S$ 
and lead to a vanishing
conductance. 
Furthermore, the GS matrix elements close to the diagonal behave 
differently as a function of $|M|$ for the two models:
they decrease with $|M|$ for SDFT and increase for the spin model, 
which is reflected in the conductance [Fig.~\ref{step1}(a)] for $V_b < 1.485$ V.
Based on Figs.~\ref{plot_overlaps}(a-c), 
we expect the giant-spin model 
to agree better with the SDFT calculation for transitions involving 
the first excited-state spin multiplet of the anion.
Indeed, Fig.~\ref{step1}(a) shows that for bias voltages 
$V_{b} \ge 1.485$ V
the two models yield qualitatively the same conductance.
The small matrix elements in Fig.~\ref{plot_overlaps}(c) 
are also the cause of the NDC seen in Fig.~\ref{dc2D}(b)
along the line separating the transport
regions $(0,+1)$ and $(-1,0,+1)$: 
when $V_{b}$ becomes large enough 
to access the anion GS multiplet,
the system remains
trapped in these states due to their small connection to the neutral states. 
Thus, the current decreases.
 
In conclusion, we presented a microscopic study of the GS properties and low-energy 
spin excitations for the neutral and charged Mn$_{12}$ molecular magnet, based on SDFT. 
Resonances in the tunneling conductance
are governed both by spin and spatial selection rules.
The latter ones
play a key role
in determining the relative contribution to transport of various spin multiplets,
and can lead to NDC. 
The orbital properties of the spin states provided by SDFT are essential to
build a correct effective spin model and interpret the transport experiments.

This work was supported by
the Faculty of Natural Sciences at Kalmar University and the Swedish
Research Council under Grant No. 621-2007-5019.
M.R.P. thanks the DOD HPCMO for computational resources.

{\it Note added} \--- After this manuscript was submitted, 
another Letter \cite{barraza_lopez2009} related to our work appeared.






\end{document}